\documentclass[11pt,twocolumn]{article}

\usepackage[english]{babel}
\usepackage[utf8]{inputenc}
\usepackage{amsmath}
\usepackage{graphicx}
\usepackage[colorinlistoftodos]{todonotes}

\title{\textbf{FilterPlus: A real-time content filtering extension for Google Chrome}}

\author{Bofin Babu, Mohan Kumar \\ Department of Computer Science \\ BITS-Pilani Hyderabad Campus \\ \{h2013313085, h2013313083\}@hyderabad.bits-pilani.ac.in}

\date{}

\begin{document}
\maketitle

\begin{abstract}
\textit{Content filtering in web browsers is a tedious process for most of the people, because of several reasons. By blocking JavaScript, Cookies and Popups, end users can ensure maximum protection from browser based attacks and vulnerabilities. In order to accomplish this, we built an extension for Google Chrome which allows users to have easy control over what they wish to recievce froma web page. We also build this extension in such a way that it remembers the choice of options made by the user for every URLs, thereby letting user’s create rules for  websites they visit.}
\end{abstract}

\section{Introduction}
Extensions are programs written to enhance the functionality of web browsers. They provide developers a  platform to build browser based applications and helps users to improve their web browsing experience. In Google Chrome web browser, extension are quite popular [1]. Among the installed extensions in Chrome, a good percentage of share goes to Popup blockers[2] and JavaScript blockers. Although many security extentions are avilable online for users, most of them lack the essential features to let user's control what they want to see from a web page. \par 
Chrome, being the most popular web browser [3], has added several security features in the recent years. Yet, majority of the users are still under the risk of attacks through it. During the past years the numbers of web based attacks showed a huge increase [4]. \par 
\begin{figure}
\centering
\includegraphics[width=0.3\textwidth]{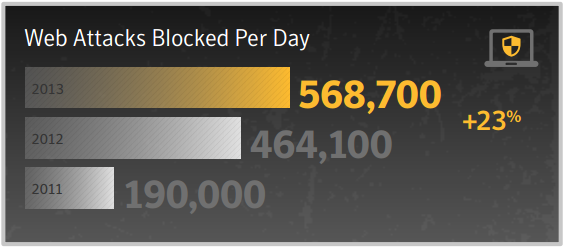}
\caption{\label{fig:symantec}Web based attacks blocked per day during the years 2011, 2012 and 2013 by Norton Internet Security software[4].}
\end{figure}
\par 
Since browsers are the primary source of internet traffic, the above statistics leads us to a conclusion that, if can implement an effective methos of blocking web contents, we can reduce a considerable amount of attacks happening all over the world.  This was the primary motivation of our project, to develop a security extension which would allow users to choose what they want receive from a url. \par 
In this project, we develop a content filtering extension named "FilterPlus" targeting Chrome browser. We included features for controlling cookies, images, popups JavaScript and notifications in our extensions.  We also made in such a way that the blocking rule created for a particular website will be remembered by the browser applies automatically whenever the user revisits the same. \par

\section{Basics of Chrome Extensions}

Google chrome extensions are basically built using HTML,CSS and JavaScript. The essential part of every Chrome extension is a manifest.json file.  This  manifest file is nothing more than a JSON-formatted table of contents, containing properties of the extention. At a high level, it is used to specify Chrome what the extension is going to do, and what permissions it requires in order to do those things. \par 
The first line of the manifest file specifies the manifest version. Since the manifest version 1 was depreciated in Chrome 18, developers are currently recommended to specify \textit{'manifest\_version':2} in the manifest file. The line follows includes three parameters:“name”, “description” and “version”, which specifies the name, description and current version of the extension. The next parameter is “permissions”, which basically use \textit{chrome.permissions} API to request declared optional permissions at run time. The next line follows “browser\_action”, which alows browser actions to put icons in the main Google Chrome toolbar, to the right of the address bar. Normally it contains “default\_icon” parameter which specifies the icon to be displayed in the browser and a “default\_popup” parameter which specifies the popup window, which will be displayed when the user clicks the icon. \par
The html file corresponding to “defult\_popup” contains  HTML code for the popup. In most of the cases it will also link to a CSS(Cascading Style Sheets) file which describes the look and formatting of the HTML document. For extension that perform some specific tasks rather than merely displaying a markup content, will also need to be linked with a JavaScript file to perform the required task. This JavaScript will let the developer  process the web content and to make API calls to the browser core and/or to some external applications -  depending on the purpose. A recent standard adopted by Google Chrome strictly probibite the addition of JavaScript file inside the HTML documents[5], there by making the need for a separate JavaScript file liked to the popup HTML document.

\section{Design Considerations}

Extensions are platform dependent. An extension written for Chrome will not work on other browsers, say Firefox or Safari, unless it is being rewritten to support them. Special care has to be taken when developing Chrome extensions since Chrome incorporates many security features and implements privilege escalation. Three security criteria included on Chrome are Safe Browsing, Sand Boxing and Auto-updates. Safe browsing feature gives warning message to the user when he/she is trying to visits a potentially malicious webpage. The Sand Box adds an additional layer of protection to the browser by protecting against malicious web pages that try to leave programs on the user’s computer, monitor his/her web activities or steal any form of private information from the hard drive [6]. The Auto-update feature enables the browser to check for updates periodically to make sure that it’s all ways up to date to ensure better protection. \par 
Chrome extensions are also divided into three components, each with progressively more privileges and less exposure to malicious web content. \par 
\begin{figure}
\centering
\includegraphics[width=0.3\textwidth]{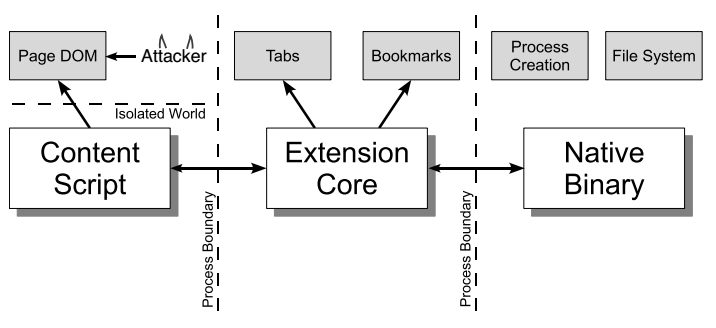}
\caption{\label{fig:privilege}Extensions are divided into three components: content scripts, an extension core and a native binary [7]
}
\end{figure}
\par 

The content script has direct access to the DOM(Document Object Model) of a web page and is exposed to potentially malicious input. The extension core, which has the bulk of the extension privileges, interact with the web content via XMLHTTPRequest and content script. Native Binary is an optional part of extensions that needs arbitrary file access on the host machine. By separating these three components Chrome achieves a great amount of security from vulnerabilities affecting through its extensions. \par

\section{Design Principles}
We adapted four design criteria and secure usability principles from [8] for the design of this extension. \par
\begin{enumerate}
\item 	\textbf{Provide branding, prevent spoofing}: Every extension uploaded into the Chrome web Store is assigned a unique key pair. The extension's ID is based on a hash of the public key, thereby providing authenticity. We’ve also made a logo and a tile – “FilterPlus” to our extension so that it will provide a unique look and feel.
\item \textbf{	Effectiveness for naïve and off-guard users}: This extension is easy to understand and has a simple GUI, such that even a user with no prior technical knowledge could use it effectively.
\item \textbf{	Minimize/avoid user work}: This extension only requires minimum user effort. The user does not need to edit the settings of the browser to make desired changes. He/She can also apply the same rule, without the need for repetition whenever a previously defined URL is revisited.
\item \textbf{	Security must be usable to be used }: Users may disable the security mechanisms which are hard to use or annoying, and it won’t affect the other functions of the extension.
\end{enumerate}

\section{Final Design and Implementation}
Keeping in mind the desired design considerations and principles, we developed the extension - “FilterPlus”. The HTML,CSS and JavaScript source files are properly liked and loaded. It is then packed, uploaded and made available for the public. \par 
\begin{figure}
\centering
\includegraphics[width=0.3\textwidth]{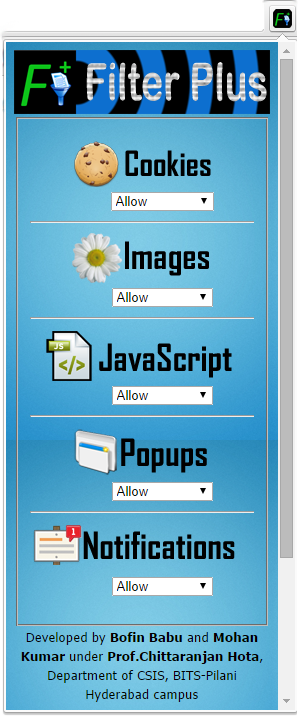}
\caption{\label{fig:extension}User interface of FilterPlus}
\end{figure}
\par
The “Cookies” module has three options in the drop down menu, namely "Allow", "Session only" and "Block". The "Allow" option enables cookies and the "Block" option disables the same for the current URL. The "session only" option allows cookies to be set only for the current session and they will be removed when a new session starts. Disabling cookies will prevent sites from storing confidential user information in the host computer. \par 
The “Images” module allows user to block images in the current URL, if they want. Web sites containing obscene images can be made safe for work(SFW) using this feature. \par
The “JavaScript” module also has two option, either to enable or disable JavaScript in the current web page. Through this features, users can make sure that no JavaScript based attacks originates from the current web page affects their system. \par 
The “Popups” module allows an option to block Popups in the current tab. Since most of the Adwares make use of popups, disabling them will protect the users from Adwares. \par
The “Notifications” has three options, either to allow, block or ask-and-allow browser notification that are displayed in the desktop. Notification can be often annoying or may contains links to third party advertisements. This can be effectively disabled using this feature. \par

\section{Conclusion}
The focus of this project was to develop a real-time content filtering extension for Google Chrome. As proposed, we have developed an extension which can filter images, popups, JavaScript codes, cookies and notifications based on user preferences. This extension can ensure user, a reasonable amount of control over what they see, using a simple GUI.  \par 

\section{References}
\begin{enumerate}
\item  Chromium Blog. "Year of Extensions", http://blog.chromium.org/2010/12/year-of-extensions.html
\item  ClarityRay Inc, "AdBlocking Report", 2012
\item Wikimedia Foundation. "Wikimedia Traffic Analysis Report – Browsers e.a", 2014, 
\item  Symantec."Internet Security Threat Report", 2014 
\item A. Barth, A.P. Felt, P.Saxena and A. Boodman, "Protecting Browsers from Extension Vulnerabilities", 2009
\item Google. "Google Chrome and Browser Security", http://tools.google.com/dlpage/res/chrome/en-GB/more/security.html
\item A. Herzberg and A. Jbara, "Security and Identification Indicators", 2006
\end{enumerate}

\end{document}